\documentclass{elsart}

\usepackage{graphicx}
\usepackage{epsfig}

\usepackage{amssymb}

\begin{document}

\begin{frontmatter}

\title{Detector Time Offset and Off-line Calibration in EAS Experiments}

\author[label11]  {H.H. He\corauthref{cor1}},
\corauth[cor1]    {Corresponding author. Tel: +86 10 88233167; Fax: +86 10 88233086} \ead{hhh@ihep.ac.cn}
\author[label3]   {P. Bernardini},
\author[label3]   {A.K. Calabrese Melcarne},
\author[label11]  {S.Z. Chen}

\address[label11] {Key Laboratory of Particle Astrophysics, Institute of High Energy Physics,
                  Chinese Academy of Sciences, Beijing 100049, Beijing, China}
\address[label3]  {Dipartimento di Fisica dell'Universit\`a del Salento and INFN, 73100 Lecce, Italy}

\begin{abstract}
In Extensive Air Shower (EAS) experiments, the primary direction
is reconstructed by the space-time pattern of secondary particles.
Thus the equalization of the transit time of signals coming from
different parts of the detector is crucial in order to get the
best angular resolution and pointing accuracy allowed by the
detector. In this paper an off-line calibration method is proposed
and studied by means of proper simulations. It allows to calibrate
the array repeatedly just using the collected data without
disturbing the standard acquisition. The calibration method is
based on the definition of a Characteristic Plane introduced to
analyze the effects of the time systematic offsets, such as the
quasi-sinusoidal modulation on azimuth angle distribution. This
calibration procedure works also when a pre-modulation on the
primary azimuthal distribution is present.
\end{abstract}

\end{frontmatter}


\section{Introduction}

In EAS experiments, the space-time information of the secondary
particles is used to reconstruct the primary direction \cite{kaska,eas,auger}. 
The space information refers to the detector unit position while the time
information is achieved usually by TDC (Time to Digital Converter).
The former is easy to measure and stable in a long period, while
the latter depends on detector conditions, cables, electronics, etc,
and usually varies with time and environment. The time offsets are
the systematic time differences between detector units, which lead
to worse angular resolution, and more seriously, to wrong reconstruction
of the primary direction. As a consequence the azimuthal distribution
is deformed according to a quasi-sinusoidal modulation \cite{Elo99}.
Thus the correction of these systematic time offsets \cite{ASgamma} is 
crucial for the primary direction reconstruction, much more when the EAS 
detector is devoted to gamma ray astronomy and the pointing accuracy is 
required in order to associate the signals with astrophysical sources.
Usually manual absolute calibration by means of a moving probe
detector is used in EAS arrays, but this method takes time and
manpower. The difficulty increases taking into account that
periodical checks are necessary to correct possible time-drift of
the detector units due to change in the operation conditions.
Furthermore the number of detector units in current EAS arrays is
getting larger and larger. As a conclusion, effective off-line
calibration procedures are greatly needed because they do not
hamper the normal data taking and can be easily repeated to
monitor the detector stability.

Here a new off-line calibration procedure is presented. It does not depend on
simulation and is very simple in the case of a uniform azimuthal distribution.
It works also when some small modulation of the azimuthal distribution is expected,
for istance due to the geomagnetic field. 
The correctness of this calibration method has been checked by means of simple
simulations both in the case of uniform and modulated azimuthal distribution.

\section{Characteristic Plane}

In EAS experiments, for an event $i$ the time $t_{ij}$ is measured on each
fired detector unit $j$, whose position ($x_j$, $y_j$) is well known.
The primary direction cosines
$l_i = \sin\theta_i \cos\phi_i$, $m_i = \sin\theta_i \sin\phi_i$
($\theta_i$ and $\phi_i$ are zenith and azimuth angles) can be reconstructed by
a least squares fit. Taking into account the time offset $\Delta_j$ typical of
the detector unit and assuming that the shower front is plane and the time-spread
due to its thickness is negligible, the plane-equation is
\begin{equation}
\label{rp} c (t_{ij} - \Delta_j - t_{0i}) = l_i x_j + m_i y_j
\end{equation}
where $c$ is the light velocity, and $t_{0i}$ is another parameter of the fit.
But the time offset $\Delta_j$ is unknown and the goal of the calibration is just to
determine it.
A traditional off-line calibration method is based on the study of the time-residuals
but their removal does not guarantee the removal of the complete offset. Therefore 
one can assume that the time offset $\Delta_j$ is the sum of two terms: the residual 
term and another unknown term. Being unaware of $\Delta_j$ the plane-equation goes like:
\begin{equation}
\label{fp} c \left(t_{ij} - t'_{0i}\right) = l'_i x_j + m'_i y_j
\end{equation}
giving the fake direction cosines
$l'_i=\sin\theta'_i \cos\phi'_i$, $m'_i=\sin\theta'_i \sin\phi'_i$.
From Eq.s~\ref{rp} and \ref{fp} and neglecting the residuals, it results:
\begin{equation}
\label{cp} \Delta_j = a_i \frac{x_j}{c} + b_i \frac{y_j}{c} + \delta_{0i}
\end{equation}
where $a_i = l'_i - l_i$, $b_i = m'_i - m_i$, and $\delta_{0i} = t'_{0i}-t_{0i}$
is an irrelevant time-shift equal for all the units. One can conclude that the
offset $\Delta_j$ is correlated with the position of the detector unit. The
quantities $a_i$, $b_i$ define a Characteristic Plane (CP) in the
($x$, $y$, $\Delta$) space, depending only on the fired unit pattern, representing the
difference between the reconstructed plane without
considering the time offset (FP: Fake Plane) and the real one (RP: Real Plane). Events
firing different sets of units have different CPs, while events firing the same set
of units have the same CP, that is the difference between the FP and RP is the same.
We define the CP of an EAS array like the average difference between FPs and RPs, i.e. the
systematic deviation between FP and RP (the pointing accuracy). The CP is fully determined
by the direction cosines
\begin{equation}
\label{eq:a} a = \left\langle l' \right\rangle -\left\langle l \right\rangle =\sin\theta_0
\cos\phi_0,
\ \ \ \ \ \ \
b = \left\langle m' \right\rangle -\left\langle m \right\rangle =\sin\theta_0
\sin\phi_0
\end{equation}
associated to the angles $\theta_0$, $\phi_0$.

\subsection{\label{sec:sinusoidal}Quasi-Sinusoidal Modulation}

If the probability density function (PDF) of the primary azimuth angle is $f(\phi|\theta)$,
one can deduce that the presence of the CP introduces a quasi-sinusoidal modulation of the
reconstructed azimuth angle distribution :
\begin{equation}
\label{f_phi0} f'(\phi'|\theta) = f(\phi|\theta)
\left[1+\frac{r}{\sqrt{1-r^2\sin^2(\phi'-\phi_0)}}\cos(\phi'-\phi_0)\right]
\end{equation}
where $r = \sin\theta_0/\sin\theta$. The PDF of the reconstructed azimuth angle is
a combination of multi-harmonics of odd orders with the the amplitude approximately
proportional to $r^{2n+1}$ ($n = 0, 1, 2...$) when $r<<1$. The time offset
does not introduce even order modulations into the reconstructed azimuth angle distribution.
When $f(\phi|\theta) = 1/2\pi$, the first harmonic becomes dominant and the PDF
of the reconstructed azimuth angle goes as
\begin{equation}
\label{f_phi1} f'(\phi'|\theta) = \frac{1}{2\pi}\left[1+r \cos(\phi'-\phi_0)\right]
\end{equation}
One can observe that the modulation parameters depend on the angles $\theta_0$ and
$\phi_0$ connected to the CP (see Eq.s~\ref{eq:a}). The phase is just $\phi_0$, while
the amplitude is proportional to $r$.
By integrating $f'(\phi'|\theta)$ over $\theta$ it results
\begin{equation}
\label{f_phi2} f'(\phi') =
\frac{1}{2{\pi}}\left[1+\sin\theta_0 \left\langle \frac{1}{\sin\theta} \right\rangle \cos(\phi'-\phi_0)\right]
\end{equation}

\begin{figure}[ht]
\begin{center}
\includegraphics*[scale=0.5]{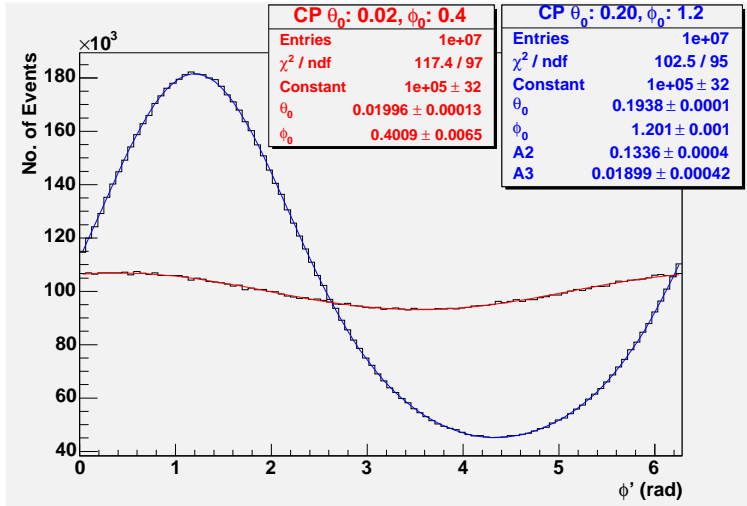}
\caption{Azimuth angle distributions fitted with harmonic functions (see the text for comments).}
\label {toymc}
\end{center}
\end{figure}

A fast Monte Carlo simulation was done to check the above conclusion. The azimuth angle was sampled
uniformly over $[0, 2\pi]$ and the zenith angle from a typical distribution modulated according to
$\cos^6\theta$ (the mode value is $\sim 22^\circ$ and $\left\langle 1/\sin\theta \right\rangle = 3.44$).
CPs with different $\theta_0$ and $\phi_0$ were assumed, subtracting $\sin\theta_0 \cos\phi_0$ and
$\sin\theta_0 \sin\phi_0$ from the original direction cosines, respectively, in order to get the new
direction cosines. Fig.~\ref{toymc} shows the reconstructed azimuth distributions for two different
CPs with $\theta_0=0.02\ rad$, $\phi_0=0.4\ rad$ and $\theta_0=0.20\ rad$, $\phi_0=1.2\ rad$, respectively.
The first distribution is well reproduced by a best-fit function like that of the Eq.~\ref{f_phi2} as
expected for small values of $\theta_0$. Also the fit parameters are in agreement with the simulation
parameters. The higher order harmonics must be taken into account in order to well reproduce the second
distribution ($A2$ and $A3$ are the amplitudes of 2nd and 3rd harmonics) because in this case $\theta_0$
and $r$ are larger.

\section{Characteristic Plane Method}

According to Eq.~\ref{rp}, if $l_i$ and $m_i$ were exactly known, then any
event can be used to relatively calibrate all the detector units hit by that
shower, while $l'_i - a$ and $m'_i - b$ can be taken as unbiased estimate
of $l_i$ and $m_i$. Therefore the time correction is determined by $a$ and $b$,
i.e. the CP of the EAS array, according to Eq.~\ref{cp}.

Suppose that the primary azimuth angle is independent on the zenith angle and distributes
uniformly, then $\left\langle l \right\rangle=0$, $\left\langle m \right\rangle=0$. Thus
$a=\left\langle l' \right\rangle-\left\langle l \right\rangle=\left\langle l' \right\rangle$,
$b=\left\langle m' \right\rangle-\left\langle m \right\rangle=\left\langle m' \right\rangle$,
which means that the CP of an EAS array can be determined by the mean values of the
reconstructed direction cosines. Then the time offsets can be calculated by means of
the off-line analysis of the collected data.

\subsection{A simple simulation as a check of the CP method \label{sec:simple}}

\begin{figure}[ht]
  \begin{center}
     \mbox{\epsfig{file=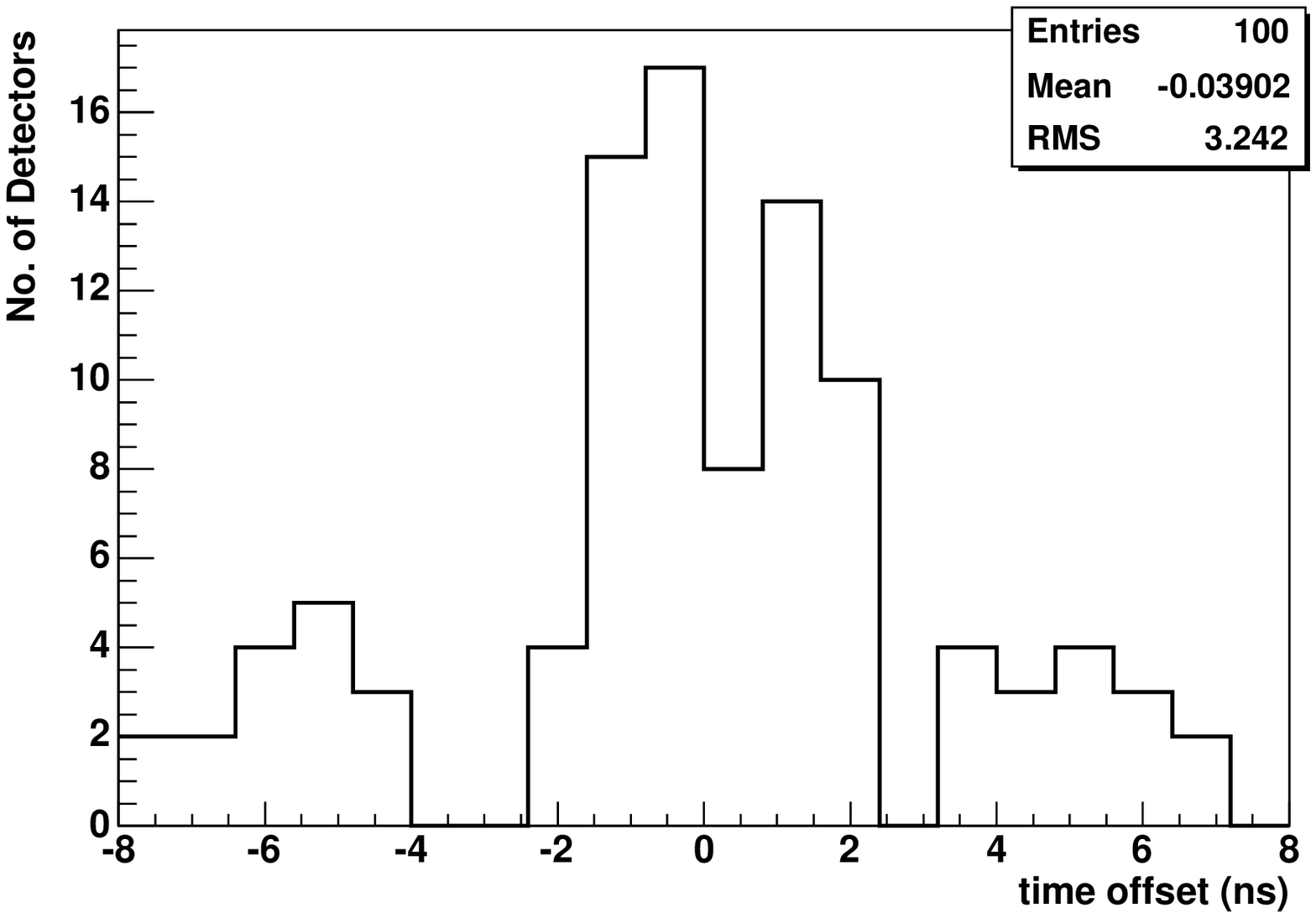,height=4.65cm}}
     \mbox{\epsfig{file=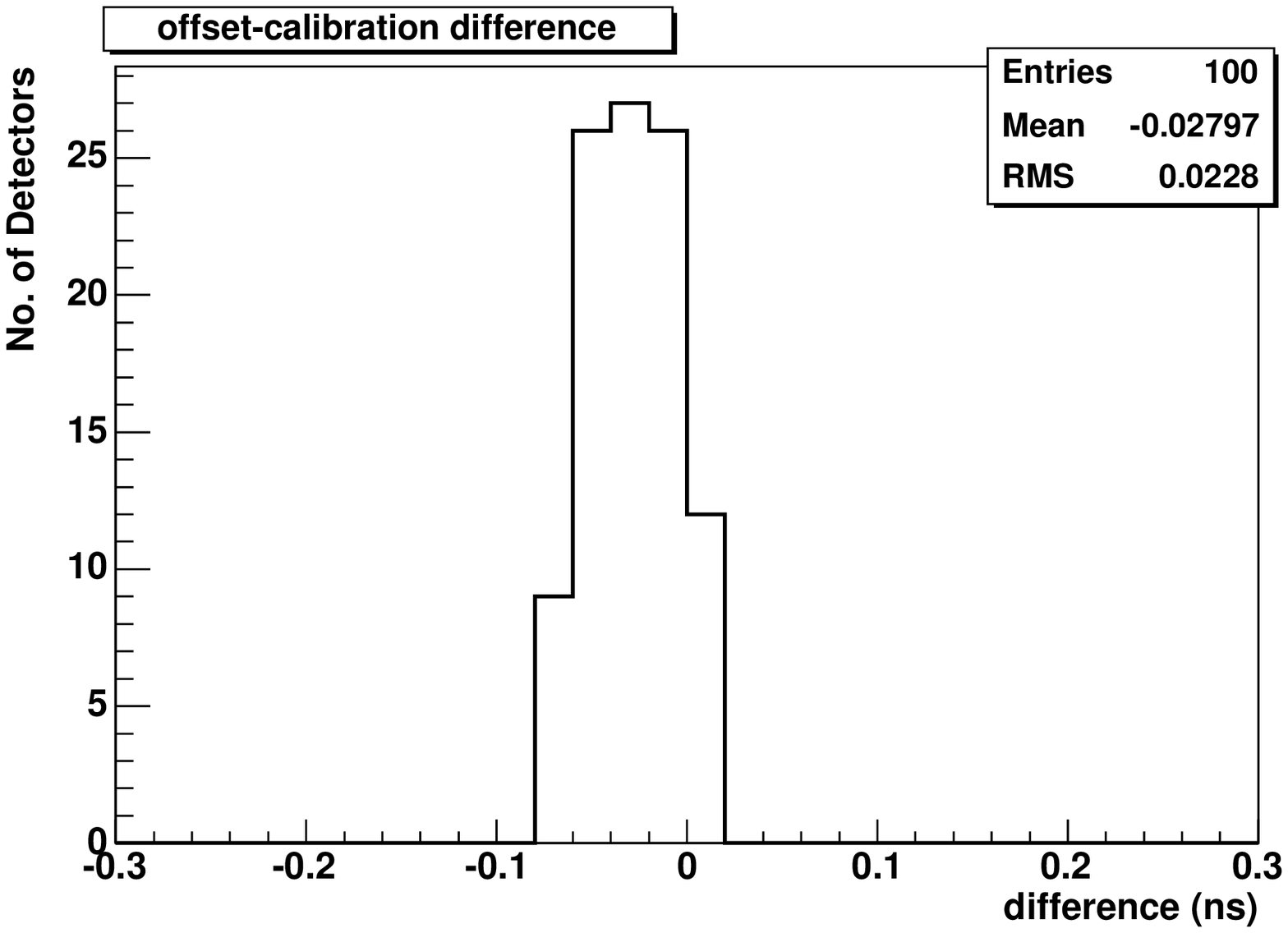,height=4.65cm}}
     \caption{First plot: systematic time offsets introduced
     in the simulation of the time measurement. Second plot:
     differences between systematic offset and calibration
     correction. \label{fig:prima}}
  \end{center}
\end{figure}

Another fast geometrical simulation was implemented in order to
check the CP method. One million of showers were extracted from
the same distributions of $\theta$ and $\phi$ used in
Sec.~\ref{sec:sinusoidal}. The arrival primary directions were
reconstructed by an array of detector units ($10 \times 10$ units
on a surface of $40 \times 40\ m^2$). The times measured by each
unit were shifted by systematic time offsets (first plot of
Fig.~\ref{fig:prima}). As a consequence the primary directions
were reconstructed with respect to a CP with $a=3.97 \times
10^{-3}$ and $b=7.34 \times 10^{-3}$ (mean values of the
reconstructed direction cosines). From the Eq.s~\ref{eq:a} it is
trivial to estimate $\theta_0 = 8.3 \times 10^{-3}\ rad$ and
$\phi_0 = 1.07\ rad$.

\begin{figure}[ht]
  \begin{center}
     \mbox{\epsfig{file=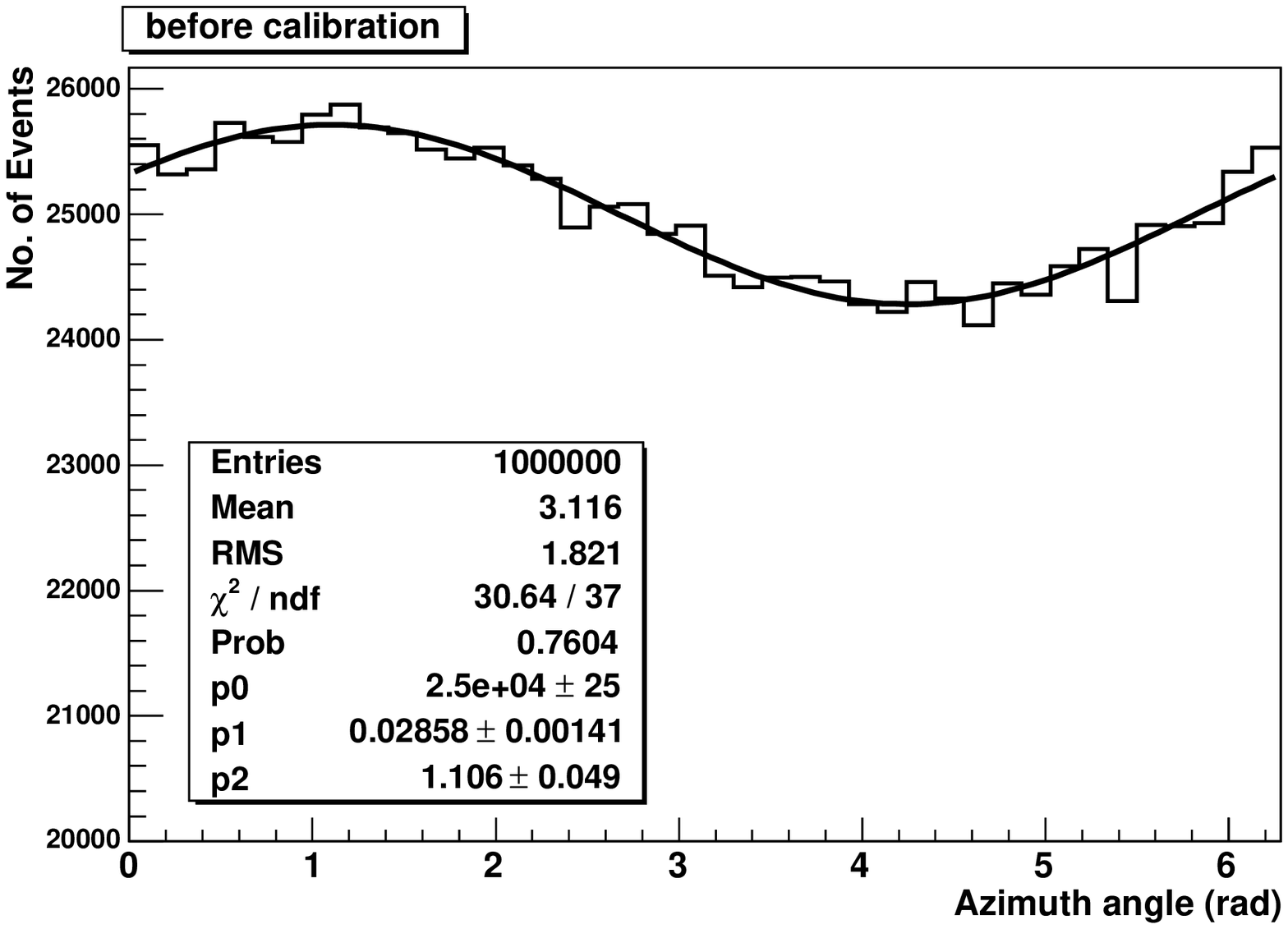,height=4.65cm}}
     \mbox{\epsfig{file=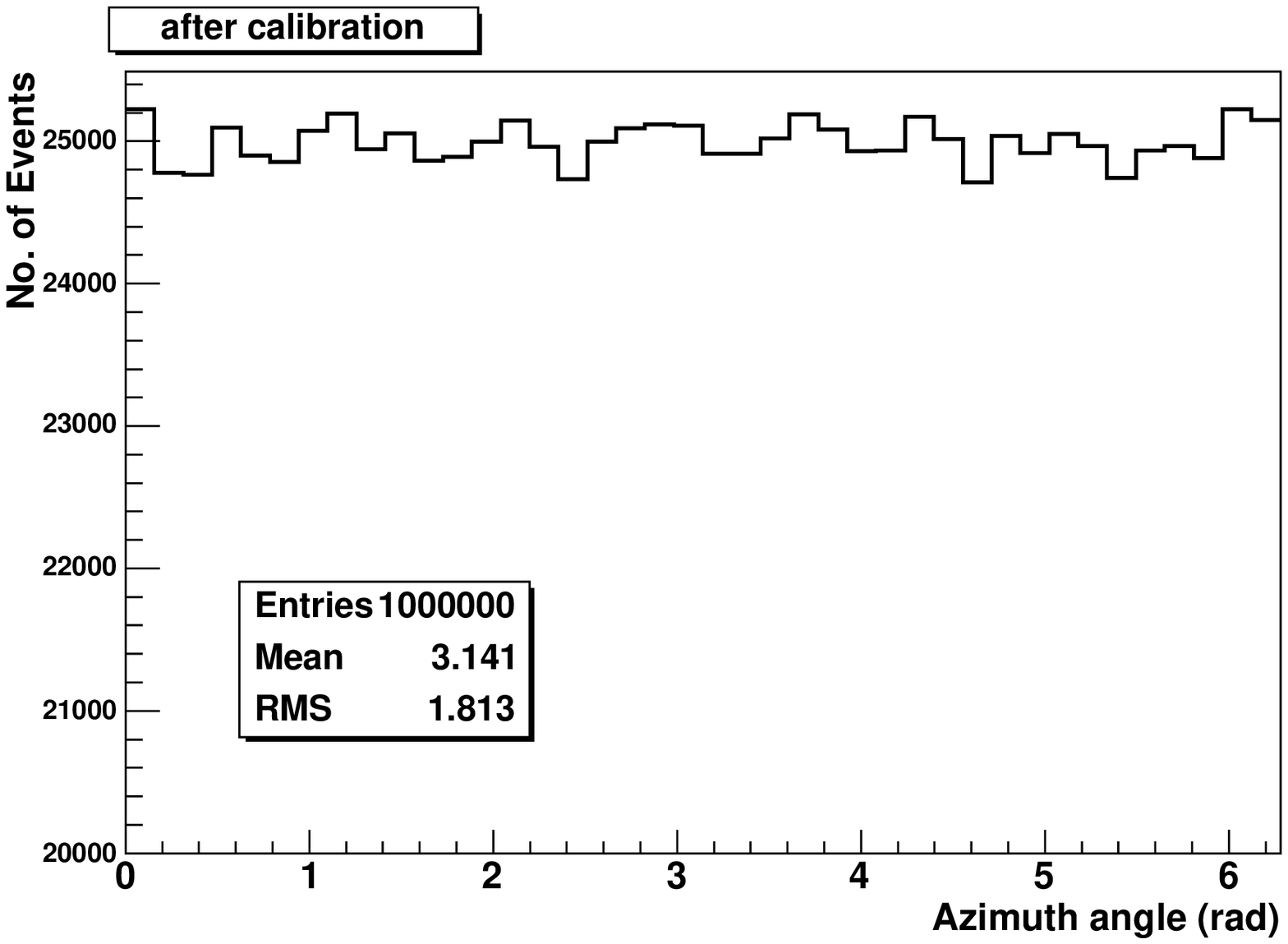,height=4.65cm}}
     \caption{Azimuth distribution (first plot: before CP calibration,
     second plot: after CP calibration). In the first plot the
     sinusoidal fit is superimposed. \label{fig:nuova}}
  \end{center}
\end{figure}

\begin{figure}[ht]
  \begin{center}
     \mbox{\epsfig{file=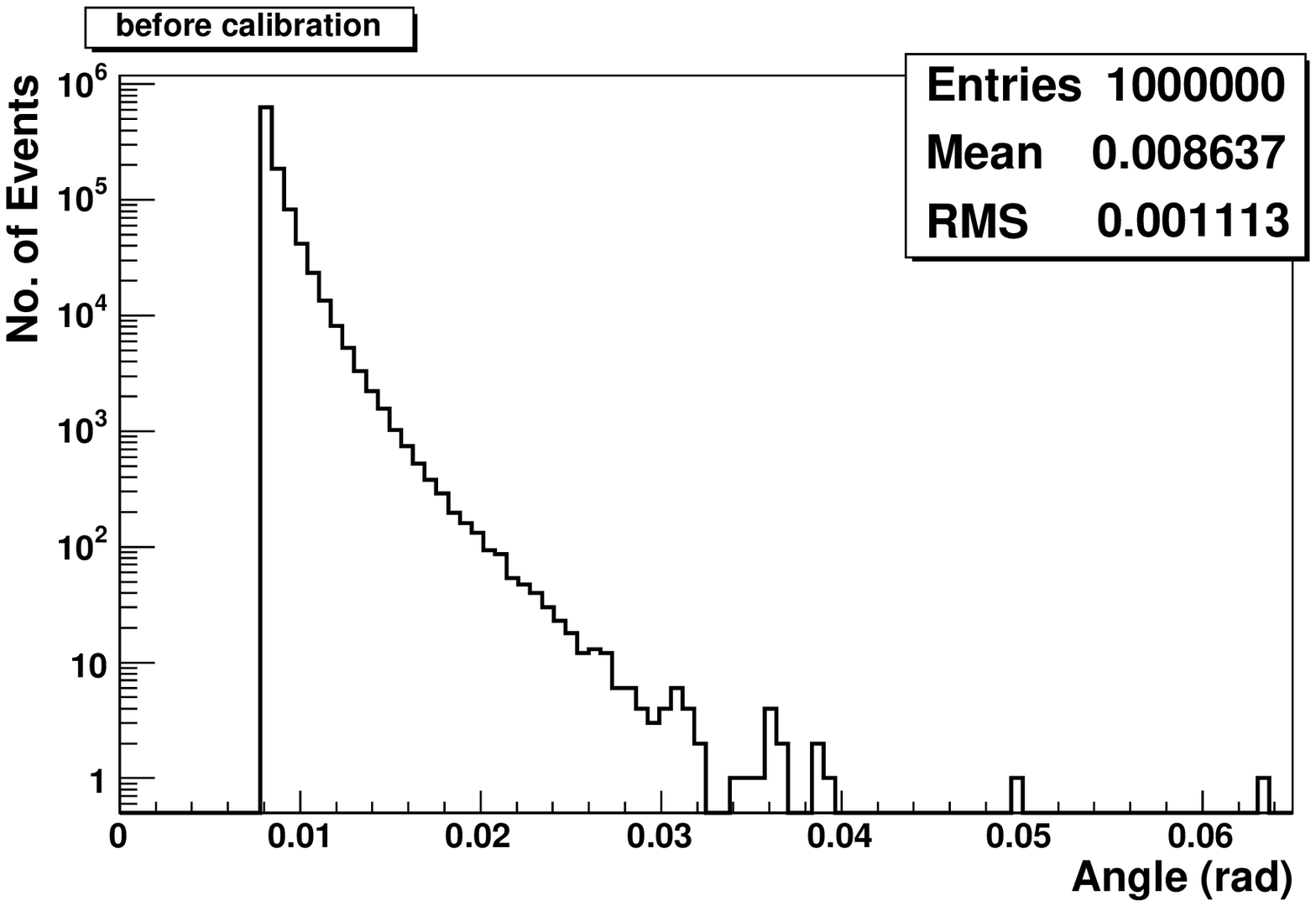,height=4.62cm}}
     \mbox{\epsfig{file=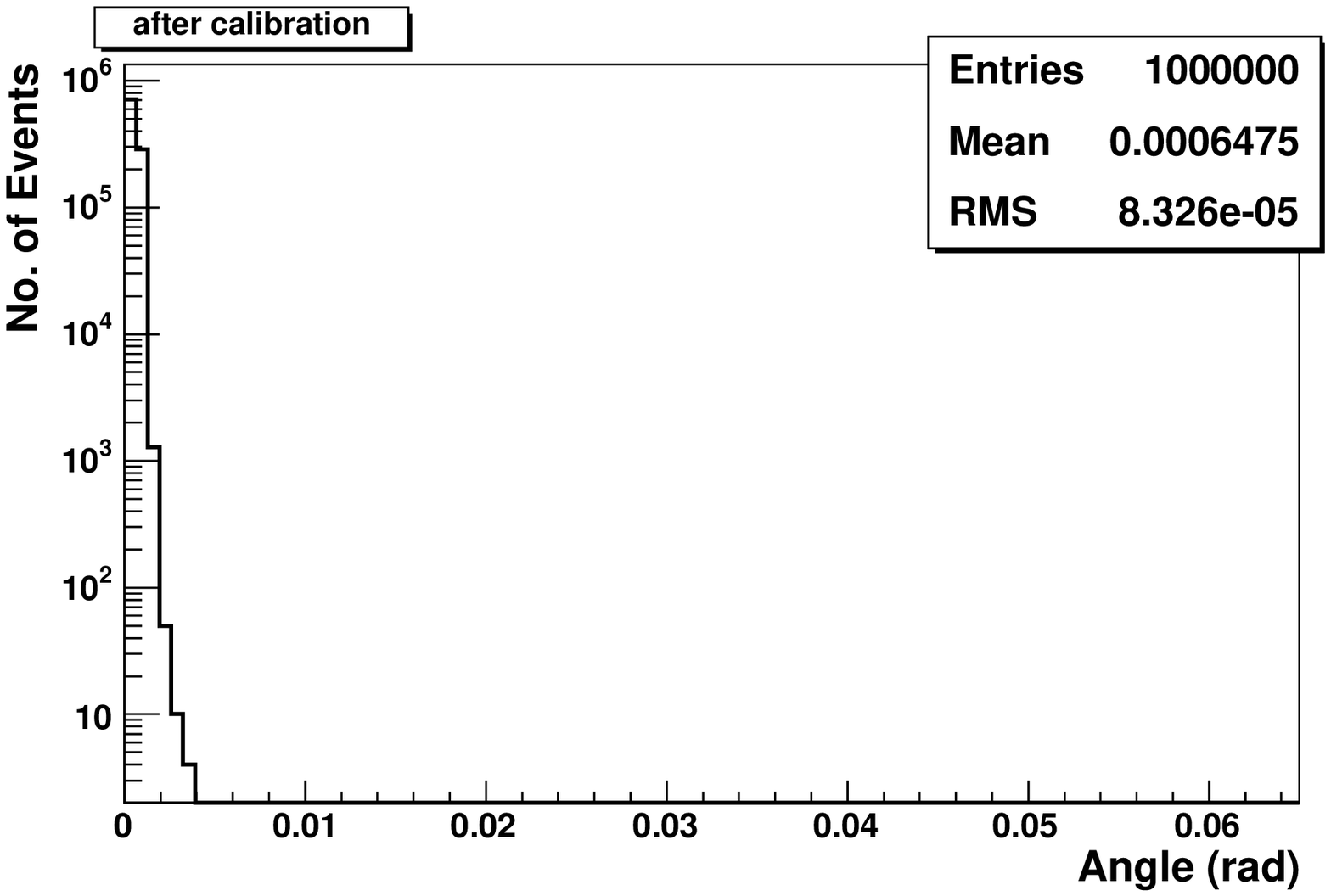,height=4.62cm}}
     \caption{Angles between "true" and reconstructed directions 
     (first plot: before CP calibration, second plot: after CP 
     calibration). \label{fig:seconda}}
  \end{center}
\end{figure}

The reconstructed azimuth distribution is fitted according to Eq.~\ref{f_phi2} (see the
first plot of Fig.~\ref{fig:nuova}). As expected the modulation coefficient $p1$ and the
phase $p2$ are compatible with $\sin\theta_0 \left\langle 1/\sin\theta \right\rangle = 0.029$
and $\phi_0$, respectively. The angles between reconstructed and "true" direction are 
shown in the first plot of Fig.~\ref{fig:seconda}.

The calibration based on the CP method allows to correct the time measurements, removing
the effect of the time offset on each detector unit. In the second plot of
Fig.~\ref{fig:prima} the offset-calibration differences are almost null and the RMS is
lower than $2.3 \times 10^{-2}\ ns$. As an effect of the CP calibration the modulation
disappears in the azimuth distribution (second plot of Fig.~\ref{fig:nuova}) and the
reconstructed directions are very close to the "true" ones (see the second plot of
Fig.~\ref{fig:seconda}). Then the validity of the CP method is fully confirmed.

\clearpage
\newpage

\vspace{0.5cm}

\vspace{0.5cm}

\clearpage
\newpage

\subsection{\label{sec:pre-modulation}Pre-modulation on the primary azimuth angle}

The assumption for the CP method is that the mean values of the primary direction
cosines are null. Generally this is not true for EAS experiments. The possible primary
anisotropy, the detection efficiency depending on the azimuth angle, the geomagnetic
effect, and so on, introduce pre-modulation into the azimuth angle distribution. Assuming
that the $\phi$-distribution is independent on $\theta$,
the pre-modulation can be described typically as:
\begin{equation}
\label{f_modulation}
f(\phi) = \frac{1}{2\pi}\left[1+\sum_{n=1}^{\infty}{g_n \cos(n\phi + \phi_n)}\right].
\end{equation}
Only $g_1 \cos(\phi+\phi_1)$ contributes to the mean values of the primary direction cosines. Therefore
they result
\begin{equation}
\label{null}
\left\langle l \right\rangle = + \frac{g_1}{2} \cos \phi_1 \left\langle \sin\theta \right\rangle,
\ \ \ \ \ \ \
\left\langle m \right\rangle = - \frac{g_1}{2} \sin \phi_1 \left\langle \sin\theta \right\rangle.
\label{llmm}
\end{equation}
The CP method annulls $\left\langle l \right\rangle$ and $\left\langle m \right\rangle$
leaving a sinusoidal modulation on the distribution of the new $\phi''$ azimuth angle.
When $g_1$ and $g_2$ are small enough and the higher order harmonics can be ignored (see
Sec.~\ref{sec:sinusoidal}) the distribution approximately is

\begin{equation}
\label{f_phi3}
  f''(\phi'')=\frac{1}{2\pi} \left[ 1 + g_1' \cos \left(\phi'' + \phi_1 \right)
                                    + g_2 \cos \left(2 \phi'' + \phi_2 \right) \right]
\end{equation}
where
\begin{equation}
\label{f_phi4} g_1' = g_1 \left[ 1 - \frac{1}{2}\left\langle
\sin\theta \right\rangle \left\langle \frac{1}{\sin\theta}
\right\rangle \right].
\end{equation}
On the basis of this result one can conclude that the calibration with the CP method does not
remove completely the pre-modulation on the primary azimuthal distribution. The $g_1$, $g_2$
amplitudes and the $\phi_1$, $\phi_2$ phases can be determined from the reconstructed azimuth
angle distribution according to Eq.s~\ref{f_phi3} and \ref{f_phi4}. Then the direction cosines
of the real CP can be determined by subtracting the pre-modulation term (Eq.s~\ref{llmm}).

\begin{figure}[ht]
\begin{center}
\includegraphics*[scale=0.5]{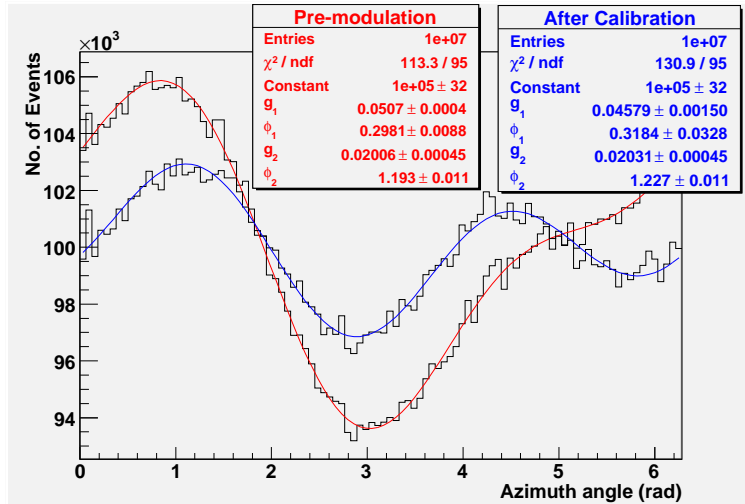}
\caption{The pre-modulation "true" azimuth angle distribution is fitted by the function
$f(\phi) = k \left[ 1 + g_1\ \cos (\phi + \phi_1) + g_2\ \cos (2\phi + \phi_2) \right]$.
The after-calibration azimuthal distribution is fitted according to Eq.s~\ref{f_phi3}
and \ref{f_phi4}. The fit-parameter values are in full agreement.}
\label{fig:prem_HHH}
\end{center}
\end{figure}

Fast simulations have been used also to check the calibration method in the case of pre-modulation with
one and two harmonics ($g_1=0.05$, $\phi_1=0.3\ rad$ and $g_2=0.02$, $\phi_2=1.2\ rad$). The
results are very similar to those of Sec.~\ref{sec:simple} confirming that the method works also
when a pre-modulation is present. In Fig.~\ref{fig:prem_HHH} the "true" azimuthal distribution
and the distribution after the first step of the calibration are shown. As expected the second
distribution is well reproduced by Eq.s~\ref{f_phi3} and \ref{f_phi4}.

\subsection{Geomagnetic Effect}

The geomagnetic field inflects the charged primaries and leads to the
well known East-West effect (with the modulation period of $\pi$ which
does not modify the mean values of the reconstructed direction cosines
and does not invalidate the CP method), while the secondary charged
particles of EAS are separated in the geomagnetic field with the lateral
distribution getting wider and flatter, thus affecting the detection
efficiency \cite{Ivanov99}. A non-vertical geomagnetic field destroys
the uniformity of the detection efficiency along the azimuth angle which
will further leads to quasi-sinusoidal modulation on the azimuth angle
distribution 
\cite{ARGOCalibrationHHH-ICRC05-3}. The geomagnetic effect on the secondaries
is typically the most significant pre-modulation (as described in
Sec.~\ref{sec:pre-modulation}) with amplitude of the order of few percent
and very slight variations with the zenith angle. This is just the case
discussed in the above section and the modulation can be determined according
to Eq.~\ref{f_phi3}, after which the time can be off-line calibrated using
the CP method.

\section{Conclusion}
The definition of the CP makes it easier to understand the effects of the detector
time offsets in EAS experiments, and makes the off-line calibration possible.
One can successfully correct the time offsets and remove the quasi-sinusoidal
azimuthal modulation (with $\phi$ depending on $\theta$).
The calibration procedure has been analytically defined and checked
by means of fast simulations. The CP calibration is very simple when the "true"
azimuthal distribution is uniform (this feature can be also achieved by selecting
special event sample). The CP method works also when a "true" pre-modulation (with
$\phi$ independent on $\theta$) of the azimuth distribution is present.
The improvement in the pointing accuracy is well shown in Fig. \ref{fig:seconda}
for the simulation of Sec.~\ref{sec:simple}. In real cases, the pointing accuracy
will depend on the detector performances and on quality and statistics of the data 
used for the CP calibration.

This method has been successfully applied to calibrate EAS detectors~\cite{ARGOCalibrationHHH-ICRC05-2}.
It has been also checked~\cite{ARGO05} by Monte Carlo full simulation and by
sampling manual calibration. The experimental results of the CP method application
will be the topic of a future paper.

\section{Acknowledgements}
We are very grateful to the people of the ARGO-YBJ Collaboration, in
particular G. Mancarella, for the helpful discussions and suggestions.
This work is supported in China by NSFC(10120130794), the Chinese
Ministry of Science and Technology, the Chinese Academy of Sciences,
the Key Laboratory of Particle Astrophysics, CAS, and in Italy by
the Istituto Nazionale di Fisica Nucleare (INFN).


\begin{thebibliography}{00}

\bibitem{kaska} T. Antoni et al., Astrophys. Journal, 608 (2004): 865-871

\bibitem{eas} M. Aglietta et al., Astroparticle Physics,  3 (1994): 1-15. 
              M. Aglietta et al., Astroparticle Physics, 21 (2004): 223-240
  
\bibitem{auger} J. Abraham et al., Nucl. Instr. Meth. A, 523 (2004): 50-95

\bibitem{Elo99} A.M. El\o\ et al., Proceedings of the 26th ICRC, Salt Lake City, 
5 (1999): 328-331

\bibitem{ASgamma} M. Nishizawa et al., Nucl. Instr. Meth. A, 285 (1989): 532-539

\bibitem{Ivanov99} A.A. Ivanov et al., JETP Letters, 69 (1999): 288-292

\bibitem{ARGOCalibrationHHH-ICRC05-3} H.H. He et al., Proceedings of the 29th ICRC, 
Pune, 6 (2005): 5-8

\bibitem{ARGOCalibrationHHH-ICRC05-2} P. Bernardini et al. for the ARGO-YBJ 
Collaboration, Proceedings of the 29th ICRC, Pune, 5 (2005): 147-150

\bibitem{ARGO05} Z. Cao et al. for the ARGO-YBJ Collaboration, Proceedings of the 
29th ICRC, Pune, 5 (2005): 299-302

\end{thebibliography}
\end{document}